\RequirePackage{fixltx2e}
\documentclass[aip,jcp,reprint,showpacs,singlecolumn]{revtex4-1}

\usepackage{dcolumn}
\usepackage{placeins}
\usepackage{epsf}
\usepackage{subfigure}
\usepackage{epsfig,amsopn}
\usepackage{amssymb,graphicx}
\usepackage{amsmath,amsfonts}
\usepackage{graphicx}
\usepackage{dcolumn}
\usepackage{bm}
\usepackage{ifpdf}
\usepackage{natbib}
\usepackage{wasysym}
\usepackage[byname]{smartref}
\usepackage{color}
\usepackage{lipsum}
\usepackage{nicefrac}
\usepackage[breaklinks, colorlinks, linkcolor=blue, citecolor=blue, urlcolor=blue]{hyperref}
\usepackage[table]{xcolor}
\usepackage{multirow}
\usepackage{braket}
\usepackage{float}
\usepackage{comment}
\usepackage{enumerate}
\usepackage{enumerate}
\usepackage{tabularx}
\usepackage[table]{xcolor}

\newcommand{\beq}{\begin{equation}}
\newcommand{\eeq}{\end{equation}}

\newcommand{\bea}{\begin{eqnarray}}
\newcommand{\eea}{\end{eqnarray}}

\begin{document}

\title{Thermopower of molecular junctions: Tunneling to hopping crossover in DNA}

\author{Roman Korol}
\affiliation{Chemical Physics Theory Group, Department of Chemistry, University of Toronto,
80 St. George Street Toronto, Ontario, Canada M5S 3H6}

\author{Michael Kilgour}
\affiliation{Chemical Physics Theory Group, Department of Chemistry, University of Toronto,
80 St. George Street Toronto, Ontario, Canada M5S 3H6}

\author{Dvira Segal}
\email{dsegal@chem.utoronto.ca}
\affiliation{Chemical Physics Theory Group, Department of Chemistry, University of Toronto,
80 St. George Street Toronto, Ontario, Canada M5S 3H6}

\date{\today}
\begin{abstract}
We study the electrical conductance $G$ and the thermopower $S$ of
single-molecule junctions, and reveal signatures of different transport mechanisms:
off-resonant tunneling, on-resonant coherent (ballistic) motion, and multi-step  hopping.
These mechanisms are identified by studying the behavior of 
$G$ and $S$ while varying molecular length and temperature.
Based on a simple one-dimensional model for molecular junctions, we derive approximate expressions for the 
thermopower in these different regimes. 
Analytical results are compared to numerical simulations, performed
using a variant of B\"uttiker's probe technique, the so-called voltage-temperature probe, which allows us to
phenomenologically introduce environmentally-induced  
elastic and inelastic electron scattering effects, while applying both voltage and temperature biases 
across the junction.
We further simulate the thermopower of GC-rich DNA molecules with mediating 
A:T blocks, and
manifest the tunneling-to-hopping crossover in both the electrical conductance and the thermopower, 
in accord with measurements by Y. Li \textit{et al.}, Nature Comm. 7, 11294 (2016). 
\end{abstract}

\maketitle

\section{Introduction}
\label{Sintro}

Complementing electrical conductance measurements,
the thermoelectric effect can serve as an accurate probe to investigate 
the transport characteristics of molecular junctions \cite{SegalmanRev,AgraitRev}.
Specifically, measurements of the thermopower  of single organic molecules reveal
whether electrons or holes are the primary charge carriers \cite{Reddy}.
The thermopower can further reveal relative orbital level-alignment \cite{Segalman1,Segalman2,Reddy12}, 
and pinpoint transport mechanisms, in support of electrical conductance measurements 
\cite{latha12,latha13,Reddy14,Reddy16,Tao,Tao16}. 

Thermoelectric devices are attractive for applications
requiring heat to work conversion without moving parts, 
a task for which nanoscale systems are particularly suitable due to their sharp energy level structure,
key for achieving high thermoelectric efficiency \cite{Sofo,Casati}.
The thermoelectric figure-of-merit of molecular junctions, $ZT$, is defined as
$ZT=GS^2T/\kappa$, with $G$ as the electrical conductance, $S$ the thermopower, and $\kappa$ the thermal conductance,
comprising both electron and phonon contributions. 
Understanding the behavior of $G$ and $S$ is therefore crucial for applications requiring enhanced 
heat to work conversion efficiency.

Measurements of the charge transfer rate and electrical conductance of single-molecule junctions and 
self assembled monolayers have revealed three primary limiting mechanisms \cite{nitzan,scheer}:
phase-coherent off-resonance ``deep'' tunneling (superexchange), 
coherent on-resonance (ballistic) conduction, and sequential incoherent hopping. 
In the deep tunneling regime the electrical conductance decreases exponentially with distance, 
becoming insignificant in long molecules.
Further, in this regime the conductance is independent of the temperature of the leads \cite{nitzan,scheer}. 

In contrast, ballistic conduction is largely insensitive to molecular length,
and is thermally activated according to the Fermi distribution of the metal electrodes \cite{Zant,Nichols}.  
If transport charges interact with ``environmental" (internal of external) degrees of freedom such as phonons,
long-range electron transfer is dominated by thermally activated incoherent hopping processes.
Such a multi-step hopping mechanism is characterized by a linear enhancement of resistance with molecular length.

The transition from off-resonant tunneling to hopping, or alternatively, to ballistic dynamics,
has been resolved experimentally by studying the distance and temperature dependence 
of the electrical conductance.
Such experiments were conducted within a wide array of organic molecules, see e.g. 
Refs. \cite{Zant,Nichols,Frisbie1,Frisbie2}, and
biomolecules, e.g. \cite{Barton10,Cahen}, revealing a consistent picture, as described above.
%
In contrast, experiments of the thermoelectric effect in single organic molecules have been primarily limited to
situations in which off-resonant tunneling was the primary transport mechanism  \cite{Reddy,AgraitRev}.
In this regime, the Fermi energy is located deep in the gap between the HOMO and the LUMO.
Landauer theory then predicts that as the electrical conductance $G$ decays exponentially 
with molecular length, the thermopower $S$ should see a linear enhancement \cite{AgraitRev}.
As well, in this deep tunneling regime 
$S$ should scale linearly with the leads' temperature.
These predictions were verified experimentally by several groups \cite{latha13,Reddy16,AgraitRev}.

While the behavior of the thermopower has thus far been assessed primarily in the deep tunneling limit,
a recent study of DNA molecules reported the crossover from tunneling to hopping behavior, 
observed simultaneously through the electrical conductance and the thermopower \cite{Tao16}.
Several DNA sequences were evaluated in this work:
In alternating (GC)$_n$ sequences, site-to-site hopping is expected to be the dominant transport mechanism,
with each purine base serving as a hopping site for holes. These sequences manifested
an ohmic behavior, with the resistance growing linearly,
and the thermopower decaying weakly-monotonically with length $n$.
In contrast, GC-rich sequences with mediating A:T blocks, 
acting as a tunneling barrier, were demonstrated 
to support superexchange behavior for short molecules,
with the conductance decreasing exponentially with the barrier width. 
Here, A, G, C and T are the adenine, guanine, cytosine and thymine bases, respectively.
By extending the A:T block, the full crossover from tunneling to hopping was
realized in the conductance. At the same time, the thermopower exhibited non-monotonic behavior:
It was linearly enhanced in the deep tunneling regime (as expected), but when 
hopping conductance was argued to dominate, it dropped down below the superexchange 
value, and then was mostly insensitive to length.
This intriguing behavior calls for theoretical work.

What theoretical-computational approaches are available to simulate $G$ and $S$, which can 
explore the full tunneling to hopping crossover in molecular junctions?
Focusing on works reporting on both $G$ and $S$, we note that
most computational studies of linear-response transport coefficients in single molecules 
are limited to the coherent limit 
\cite{Dubirev}, using the Landauer formula
with parameters derived from first principle calculations
\cite{Datta,Cuevas08, Stafford09,Stafford10, Neaton, Pauli12, Pauli15, Markussen,Gemma14,Gemma15,Lambert16},
or other coherent-transport approaches \cite{Dubi09}.
Effects of vibrations on the thermopower were assessed 
using scattering approaches \cite{Chen11,Chen12}, or perturbatively using
non-equilibrium Green's function (NEGF) \cite{Galperin,Ora,perroni,Agarwalla15,AgarwallaBeil,Brandbyge} 
or quantum master equation (QME) methods \cite{Koch,Kamil,Crystal,Agarwalla15,AgarwallaBeil}.
However, given the computational cost, such treatments are limited to describe rather small systems
with few (1-2) molecular electronic orbitals and 1-2 primary vibrational modes. 
Thus, while scattering methods, the NEGF technique, or QME tools 
can be used to identify vibrationally-assisted transport effects,
they cannot be feasibly employed to reveal the full crossover from tunneling dynamics to multi-step hopping as induced by
electron-vibration interaction processes.

The Redfield equation of motion, which does not explicitly include particular molecular vibrations,
can be used to describe charge transport in extended-interacting systems \cite{nitzan}.
In this method, the information from the thermal environment is encapsulated within Fourier transforms 
of bath correlation functions which can be generated to mimic thermalized vibrational effects.
Based on a (phenomenological) generalization of the Landauer equation to include inelastic effects,
the Redfield approach was used in Ref. \cite{SegalRed} to uncover the full tunneling-to-hopping crossover 
concurrently in the behavior of $G$ and $S$. This approach however relies on several significant assumptions:
(i)  The inelastic transmission probability, which 
allows for energy exchange with the environment, was conjectured to exist.
The Landauer framework was then assumed to hold, predicated on the existence of 
single quasi-particle scattering states.
(ii) Certain voltage and temperature profiles were presumed to  develop across the junction. 
Also, as we discuss after Eq. (\ref{eq:SH}), the ballistic contribution was incorrectly identified
as hopping conduction in some places in Ref.  \cite{SegalRed}. 
It is thus necessary to develop a new framework, based on more solid grounds,
for simulating the thermopower in systems affected by environmental interactions.

The Landauer-B\"uttiker probe (LBP) method offers an alternative route for introducing environmental
effects into charge transport dynamics at low cost. The method is particularly appealing since
it allows the study of a broad range of systems, from single-atom junctions up to the thermodynamic limit.
In this technique, incoherent elastic and inelastic scattering effects are included in a phenomenological manner,
by augmenting the non-interacting electronic Hamiltonian with probe terminals through which electrons lose their phase memory 
and (possibly) exchange energy with environmental degrees of freedom \cite{Buttiker1,Buttiker2}.
While the technique was originally introduced to study decoherence effects in mesoscopic devices, 
it was recently applied to explore electronic
conduction in organic and biological molecular junctions \cite{Nozaki1,Nozaki2,Chen-Ratner,WaldeckF}, as well as
anharmonic effects in (purely) phononic quantum conduction \cite{SC2,SCMalay,SCnano}.
Particularly, in Ref. \cite{Kilgour1} we demonstrated that the LBP
method can capture different transport regimes in molecular wires: tunneling conduction,
ballistic motion, and incoherent hopping. In Refs. \cite{Kilgour2,Kilgour3}, we further used the LBP technique
to simulate high-bias voltage effects, specifically the role of environmental interactions
on the diode operation. More recently, we demonstrated that the LBP method can uncover
an intermediate quantum coherent-incoherent transport regime in DNA junctions \cite{William}.

In this work, our objective is to study the thermoelectric effect in molecular junctions
in situations when different transport mechanisms play a role: superexchange, phase coherent 
ballistic motion, and  multi-step hopping.
We perform our simulations using two variants of the probe technique: 
voltage probe (VP), which hands over the electrical conductance,
and voltage-temperature probe (VTP), which generalizes the VP method by including both voltage and temperature biases,
allowing us to simulate the thermopower of a junction \cite{Jacquet09,Dhar07,Salil,Stafford16}.
Our specific objectives are:
(i) Develop analytic expressions for the thermopower in molecular junctions covering 
different transport regimes (superexchange, ballistic tunneling, hopping).
(ii) Understand the transition between the different mechanisms based on numerical simulations. 
(iii) Use simulations to explore the recently reported tunneling-to-hooping crossover in DNA 
molecules as manifested in the thermopower \cite{Tao16}.

Below we illustrate that the three transport regimes can be identified based on the 
combined length and temperature dependence of $G$ and $S$.
However, while $G$ can be suppressed by orders of magnitude as we increase molecular length, 
changes in $S$ are rather modest over the whole range. 
Another interesting observation concerns the temperature dependence of the thermopower, 
which is identical (linear) in the tunneling and hopping regimes, 
while in the ballistic regime we find that $S\propto 1/T$. 
Finally, our LBP simulations of charge transport in DNA molecules reproduce the tunneling-to-hopping turnover
in qualitative agreement with measurements \cite{Tao16}---once we suppress ballistic current.

The paper is organized as follows. In Sec. \ref{Method}, we present the voltage-temperature probe
technique. In Sec. \ref{1D},
we introduce a simple 1-dimensional (1D) model for charge conduction in linear molecules,
develop analytic expressions for the thermopower in different regimes, and support these expressions with
numerical simulations.
Simulations of double-stranded DNA junctions with A:T blocks forming tunneling barriers are included in Sec. \ref{DNA}.
We conclude in Sec. \ref{Summ}.

\begin{figure}[htbp] 
\vspace{-2mm} \hspace{-7mm}
{\hbox{\epsfxsize=88mm \epsffile{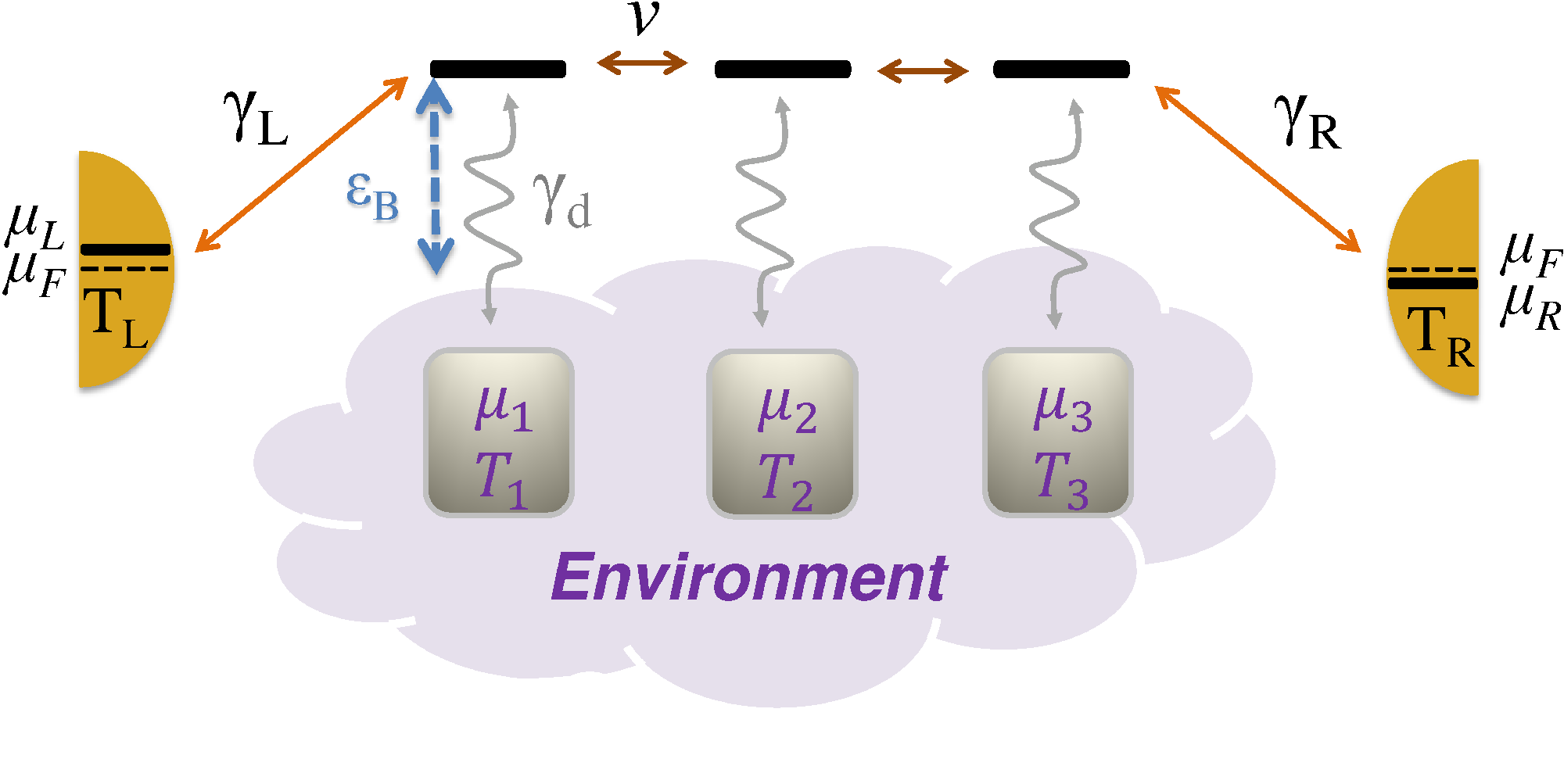}}}
\caption{Scheme of the 1D molecular junction model examined in this work for $N=3$. The conductance and the 
thermopower are calculated using the LBP method, mimicking environmental effects in the junction with probes. 
In the thermopower calculation, the inner reservoirs' 
chemical potentials and temperatures are self-consistently determined so as 
to satisfy zero charge and heat dissipation conditions.
Full-line arrows depict tunneling events in the molecule. Curly arrows describe
transitions between molecular states and the  probes.}
\label{Figs1}
\end{figure}

\section{Method: Voltage-temperature probe}
\label{Method}

The LBP method introduces environmental effects including 
decoherence, energy exchange, and dissipation into charge transport calculations.
To compute the electrical conductance, we use the voltage probe as implemented in Ref. \cite{Kilgour1}.
To simulate a thermal gradient and compute the thermopower, 
however, we have also to self-consistently set the temperature of the thermal 
environment, which we compute via the voltage-temperature probe method \cite{Jacquet09, Dhar07}.
We begin with a general presentation of the VTP technique.

The molecular structure is described by a tight-binding Hamiltonian with $N$ sites.
The Hamiltonian includes the static energies of the localized molecular orbitals 
and charge transfer integrals between molecular orbitals centered on different sites. 
Each molecular site is coupled to a probe reservoir $j=1,2,...N$, which can exchange particles and energy with the  
molecular system. 
As well, we assume that the left and right end groups are coupled to physical reservoirs (electrodes) $\nu=L,R$,
where we impose the boundary conditions: chemical potentials $\mu_{L,R} $ and temperature $T_{L,R}$.  
Below we use the index $\alpha$ to identify
all leads: the two metal electrodes $\nu=L,R$ and the $j=1,2,..,N$ probes, acting on each site. 
The total-net charge current, leaving the $L$ contact, is written as (per spin specie) 
\bea
I_L=\frac{e}{h}\sum_{\alpha} \int_{-\infty}^{\infty}
\mathcal T_{L,\alpha}(\epsilon) \left[f_L(\epsilon)-f_{\alpha}(\epsilon)\right]d\epsilon.
\label{eq:currL}
\eea
Here, $f_{\alpha}(\epsilon)=[e^{\beta_{\alpha}(\epsilon-\mu_{\alpha})}+1]^{-1}$ is 
the Fermi function in the electrodes, given in terms of the
inverse temperature $k_BT_{\alpha}=\beta_{\alpha}^{-1}$ and chemical potentials $\mu_{\alpha}$,
$k_B$ is the Boltzmann constant.
The probe parameters, $\mu_j$ and $T_j$,
are determined from the voltage-temperature probe condition, to be explained below.
The transmission functions in Eq. (\ref{eq:currL}) are obtained from
the $N\times N$-sized molecular Green's function and the hybridization matrices \cite{nitzan},
\bea
\mathcal T_{\alpha,\alpha'}(\epsilon)={\rm Tr}[ \hat \Gamma_{\alpha}(\epsilon)\hat G^r(\epsilon)\hat \Gamma_{\alpha'}(\epsilon)\hat G^a(\epsilon)].
\label{eq:trans}
\eea
The trace is performed over the molecular states. The retarded Green's function is given by
\bea
 \hat G^r(\epsilon)=[\hat I\epsilon-\hat H_{M} + i \hat \Gamma/2]^{-1},
\label{eq:Gr}
\eea
with $\hat G^a(\epsilon)=[\hat G^r(\epsilon)]^{\dagger}$, 
$\hat \Gamma=\hat \Gamma_L+\hat \Gamma_R+\sum_{j=1}^{N}\hat \Gamma_j$, 
and $\hat H_M$ the Hamiltonian of the $N$-state molecular system.
In structures considered in this work, the molecule is coupled to each metal lead through a single site,
with the left (right) lead coupled to site '1' ('$N$').
The $L,R$ hybridization matrices therefore include a single nonzero value,
\bea
[ \hat \Gamma_{L}]_{1,1}= \gamma_L, \,\,\,\,
[\hat \Gamma_{R}]_{N,N}= \gamma_R,
\eea
with $\gamma_{L,R}$ describing the metal-molecule coupling energy.
We work in the wide-band limit: We take
$\gamma_{L,R}$ as energy independent parameters and ignore energy shifts of electronic states due to the metal leads.
The hybridization matrices $\hat \Gamma_j$ describe the coupling of the $j$th probe to the respective site.
For simplicity, we assume that incoherent effects are local, uncorrelated, and uniform,
with $\gamma_d/\hbar$ as the rate constant due to environmental processes,
\bea
[ \hat \Gamma_{j}]_{j,j}= \gamma_{d},  \,\,\,\ j=1,2,..,N
\eea
%
%

We now explain the voltage-temperature probe condition which we employ in the linear response regime.
We set the chemical potentials and the temperatures at the boundaries ($\mu_L$, $T_L$) and ($\mu_R$, $T_R$).
%
%
%
At finite coupling strength $\gamma_d$, conducting electrons can tunnel back and forth
from the molecular system into the probes.
To enforce charge conservation between source ($L$) and drain ($R$), 
the charge current leaking to each probe, $I_j$, is made to nullify. 
Heat leakage $Q_j$ to the probes is also prohibited. These two 
conditions translate into $2\times N$ equations for the  probes' chemical potentials $\mu_j$ and temperatures $T_j$,
\bea
I_j&=& \frac{e}{h} \sum_{\alpha}
 \int_{-\infty}^{\infty} d\epsilon
\mathcal T_{j,\alpha}(\epsilon)
\left[f_j(\epsilon)-f_{\alpha}(\epsilon)\right] = 0
\nonumber\\
Q_j&=& \frac{1}{h} \sum_{\alpha}
\int_{-\infty}^{\infty} d\epsilon
(\epsilon-\mu_j)\mathcal T_{j,\alpha}(\epsilon)
\left[f_j(\epsilon)-f_{\alpha}(\epsilon)\right] = 0
\nonumber\\
\label{eq:VTP}
\eea
In the linear response regime (small bias and small temperature difference),  we Taylor-expand  the Fermi function,
\bea
&&f_{\alpha}(\epsilon,T_{\alpha},\mu_{\alpha})=f_{eq}(\epsilon,T_{eq},\mu_{eq})  
\nonumber\\
&&-\frac{\partial f_{eq}(\epsilon,T_{eq},\epsilon_F)}{\partial \epsilon}
\left[  \frac{\epsilon-\epsilon_{F}}{T_{eq}}(T_{\alpha}-T_{eq}) + (\mu_{\alpha}-\epsilon_F) \right].
\eea
We explicitly indicate the dependency of the Fermi function on the equilibrium temperature $T_{eq}$ and the
Fermi energy $\epsilon_F$. For convenience, below we set $\epsilon_F=0$.
The $2N$ equations in Eq. (\ref{eq:VTP}) can be organized as follows ($p=0,1$),
\begin{widetext}
\bea
\sum_{\alpha} \int_{-\infty}^{\infty} d \epsilon \mathcal T_{j,\alpha}(\epsilon) \left(-\frac{\partial f_{eq}}
{\partial \epsilon}\right)(\epsilon-\epsilon_F)^p
\left[ \frac{\epsilon-\epsilon_F}{T_{eq}}(T_j-T_{\alpha}) +(\mu_j-\mu_{\alpha})  \right] =0
\eea
\end{widetext}
The $N$ equations with $p=0$ ($p=1$) correspond to charge (heat)-current equations (\ref{eq:VTP}).
%
%
We solve this linear system and obtain the probes' chemical potentials $\mu_j$ and temperatures $T_j$.
In the next step, the set \{$\mu_j$,  $T_j$\} is used in the calculation of the net current flowing across the device.
Eq. (\ref{eq:currL}) provides the charge current leaving the $L$ terminal, which (given the probe condition)
is identical to the current arriving at the drain 
$R$. After linearizing this equation we receive
\bea
I_L&=&\frac{e}{h}\sum_{\alpha} 
\Bigg[
(\mu_L-\mu_\alpha)
\int_{-\infty}^{\infty}
\mathcal T_{L,\alpha}(\epsilon) \left(-\frac{\partial f_{eq}}{\partial \epsilon}\right) d\epsilon 
\nonumber\\
&+& (T_L-T_\alpha)  \int_{-\infty}^{\infty}
\mathcal T_{L,\alpha}(\epsilon) \left(-\frac{\partial f_{eq}}{\partial \epsilon}\right)\left(\frac{\epsilon-\epsilon_F}{T_{eq}} \right) d\epsilon \Bigg].
\nonumber\\
\label{eq:currLR}
\eea
%
The thermopower, or  the Seebeck coefficient, is defined as
\bea
S=-\frac{\Delta V}{\Delta T}\Big|_{I_L=0},
\label{eq:Seebeck}
\eea
with $\Delta \mu=e\Delta V=\mu_L-\mu_R$ and $\Delta T=T_L-T_R$.
We organize Eq. (\ref{eq:currLR}) as $I_L=L_{1,1}e\Delta V + L_{1,2}k_B\Delta T$ and write the thermopower,
in the language of the linear response coefficients as $S=\frac{k_B}{e}L_{1,2}/L_{1,1}$. 
To calculate the linear conductance of the junction we set $\Delta T=0$.
We then work with the voltage probe method \cite{Kilgour1} and study---in the linear response regime---the resistance of the junction $R\equiv (\Delta V)/I_L$, or its conductance $G=R^{-1}$. 

\section{1-dimensional model}
\label{1D}

In this Section we use a simple 1-dimensional (1D) model to derive approximate analytical results for $S$ 
in different transport regimes. We then perform  numerical simulations using the VP and the VTP methods
and present the electrical conductance and the thermopower hand in hand.
The tunneling-to-hopping transition is clearly exhibited in
the electrical conductance as a crossover from an exponential decay to a weak, Ohmic-like, distance dependence.
As we show below, the thermopower can also indicate on this transition.

\subsection{Hamiltonian}

We consider a simple tight-binding 1D model with nearest-neighbor electronic coupling $v_a$ and a uniform 
on-site energy $\epsilon_B$,
\bea
\hat H_M=\sum_{j=1}^{N} \epsilon_B \hat c_{j}^{\dagger}\hat c_j +
\sum_{j=1}^{N-1}v_a\left( \hat c_{j}^{\dagger}\hat c_{j+1}  +h.c. \right).
\label{eq:HA}
\eea
Eq. (\ref{eq:HA}) serves as the molecular Hamiltonian $\hat H_M$ in Eq. (\ref{eq:Gr}).
To capture the tunneling-to-hopping crossover we consider chains with $N=2-10$ sites
with parameters satisfying $v_a/\epsilon_B\ll1$. Specifically, we select
$\epsilon_B=0.5$ eV and $v_a=0.05$ eV.
The metal-molecule hybridization $\gamma_{L,R}$ is taken in the range 0.05 - 0.5 eV. 
We work with a wide range of temperatures, $T_{eq}=5-300$ K,
and play with environmental effects in the range $\gamma_d=0 - 0.2$ eV.

\subsection{Analytic results}

We derive here approximate analytical expressions for the thermopower in the three relevant transport limits, 
tunneling, ballistic, and hopping.
We begin by explaining the behavior of the thermopower when environmental effects are ignored.
Based on the Landauer Formula [Eq. (\ref{eq:currLR}) without probes], we receive 
\bea
S=\left( \frac{1}{eT} \right) 
\frac{
\int d\epsilon 
\mathcal T(\epsilon) \left( -\frac{\partial f_{eq}}{\partial \epsilon}\right) (\epsilon-\epsilon_F)}
{ \int d\epsilon \mathcal T(\epsilon) \left(-\frac{\partial f_{eq}} {\partial \epsilon}\right)}.
\label{eq:SC}
\eea
Here $e$ is the charge of an electron (with a negative sign).
We now evaluate this expression in two limits.
First, in the deep tunneling regime, $|\epsilon_B-\epsilon_F|\gg v_a,\gamma_{L,R},k_BT$,
we only need to consider the behavior of the transmission function at the vicinity of the Fermi energy.
Since we are missing sharp features (resonances) in this region,
we Taylor-expanded the transmission function as
\bea
\mathcal T(\epsilon) \sim 
\mathcal T(\epsilon_F) + \frac{\partial \mathcal T(\epsilon)}{\partial \epsilon}\Bigg|_{\epsilon_F}(\epsilon-\epsilon_F).
\label{eq:TTaylor}
\eea
We plug this expansion into Eq. (\ref{eq:SC}), and note that since the derivative of the Fermi function is approximately
a delta function around the Fermi energy at low temperatures, the first contribution vanishes,
and we are left with the familiar  behavior
\bea
S&=&\frac{1}{eT} 
\frac{ \frac{\partial \mathcal T(\epsilon)}{\partial \epsilon}\Big|_{\epsilon_F}
 \int d\epsilon 
\left( -\frac{\partial f_{eq}}{\partial \epsilon}\right) (\epsilon-\epsilon_F)^2}
{\mathcal T(\epsilon_F) \int d\epsilon 
 \left(-\frac{\partial f_{eq}} {\partial \epsilon}\right)}
\nonumber\\
&=&  
\frac{\pi^2k_B^2T}{3e}
\left(\frac{  \frac{\partial \mathcal T(\epsilon)}{\partial \epsilon}\Big|_{\epsilon_F}}{ \mathcal T(\epsilon_F)} \right).
\label{eq:ST}
\eea
Note that $e$ is the charge of an electron. With this sign convention, this expression agrees
with other studies, see e.g. \cite{Reddy,Reddy12,Reddy14,Reddy16,Tao}. 
We can in particular explore the superexchange limit. In this (tunneling, $T$) case, the transmission function 
decays exponentially with length \cite{nitzan},
\bea
\mathcal T_T(\epsilon)= \left(\frac{v_a}{\epsilon-\epsilon_B}\right)^{2N} \frac{\gamma_L\gamma_R}{v_a^2},
\label{eq:Tsupex}
\eea
resulting in the conductance $G_T=G_0\mathcal T_T(\epsilon_F)$, $G_0$ is the quantum of conductance. 
Using Eq. (\ref{eq:Tsupex}) in (\ref{eq:ST}), we receive what we refer to as the 'tunneling-superexchange' 
thermopower,
\bea
S_T= \frac{\pi^2}{3} \left(\frac{k_B}{e}\right) \frac{k_BT}{\epsilon_B-\epsilon_F}  \times 2N.
\label{eq:STT}
\eea
This expression predicts the following properties---which were verified in experiments 
\cite{latha12,latha13,Reddy16}:
In the deep tunneling regime, the thermopower increases linearly with temperature and with the number of 
molecular sites $N$. As well, LUMO conduction, with $\epsilon_B$ situated above the Fermi energy,
shows a negative thermopower.

We now consider the coherent on-resonant (ballistic, $B$) case. 
In the limit of weak metal-molecule hybridization, the transmission displays sharp features which we approximate
by a Dirac delta function,
$\mathcal T_B(\epsilon)\sim A \delta(\epsilon-\epsilon_B)$, $A$ is a parameter with the dimension of energy.
Using Eq. (\ref{eq:SC}), we find that ballistic electrons organize the thermopower
\bea
S_B= \left(\frac{k_B}{e}\right) \left( \frac{\epsilon_B-\epsilon_F}{k_BT}\right).
\label{eq:SB}
\eea
Interestingly, in the coherent-resonant regime the scaling of the thermopower 
with temperature and with bridge height 
$\epsilon_B$ are opposite to trends observed in the deep tunneling limit.

We now examine situations suffering from environmental effects, $\gamma_d\neq0$.
In this case, the conductance manifests hopping (Ohmic) conduction for long enough molecules.
In the language of the probe technique, we recall that at low temperatures we can organize 
an expression for the thermopower in the form of Eq. (\ref{eq:SC}), with an {\it effective} transmission function 
which comprises the contributions of the probes \cite{Pastawski}.
We can Taylor expand this effective (probe dependent) transmission function and organize the result in 
the structure of Eq. (\ref{eq:ST}).
What is the form of this probe dependent transmission function in the hopping ($H$) regime, 
$\mathcal T_{H}(\epsilon)$?
Based on simulations, in Ref. \cite{Kilgour1} we proposed the following structure:
\bea
\mathcal T_{H}(\epsilon) \sim \frac{\gamma_d^2v_a^2}{(\epsilon_B-\epsilon)^4f(N)},
\label{eq:TH}
\eea
with the conductance $G_H\sim G_0\mathcal T_{H}(\epsilon)$.
The function $f(N)$ depends on the molecular length $N$,
approximately in a linear manner.
Substituting Eq. (\ref{eq:TH}) into Eq. (\ref{eq:ST}), we arrive at a rather
simple approximation for the thermopower, valid in the hopping regime,
\bea
S_H\sim 
\left(\frac{k_B}{e}\right) 
\frac{4\pi^2}{3} \frac{k_BT} { (\epsilon_B-\epsilon_F)}.
\label{eq:SH}
\eea
%
Surprisingly, this expression 
is almost identical to the superexchange limit,  Eq. (\ref{eq:STT})---only replacing the linear dependency 
on $2N$  with a factor 4.
%
Recall that Eq. (\ref{eq:SH}) was organized based on the D'Amato Pastawski effective transmission 
formula \cite{Pastawski},
which is limited in applicability to the low temperature regime and to off-resonance situations.
Our numerical simulations below did not rely on this approximation, though we note that
since selected parameters satisfy $|\epsilon_B-\epsilon_F|\gg v_a,k_BT$, we are essentially 
considering  low-temperature situations.

Equation (\ref{eq:SH}) is rather interesting. It predicts that when the current is dominated by 
hopping conduction,
the thermopower reaches a constant value, independent of molecular length and environmental interactions
$\gamma_d$. Furthermore, this expression predicts that  $S_H\sim S_T(N=2)$. As a result,
another concrete expectation is that at low temperatures $\epsilon_B\gg v_a,k_BT$, 
$S(N)$ should be a concave function, since the hopping thermopower lies below the coherent value.
Both of these predictions agree with measurements of DNA molecules \cite{Tao16}.

Note that in Ref. \cite{SegalRed}, the hopping conduction was argued to follow  Eq.
(\ref{eq:SB}), identical to the ballistic behavior.
We point out that Eq. (25) of Ref. \cite{SegalRed} was incorrectly attributed to hopping behavior, and in fact it
corresponds as well to ballistic motion.
Thus, to the best of our knowledge, Eq. (\ref{eq:SH}) here is the first analytical construction
for the thermopower in the hopping transport regime.

\begin{figure}[htbp]
\vspace{0mm} \hspace{-10mm}
{\hbox{\epsfxsize=97mm \epsffile{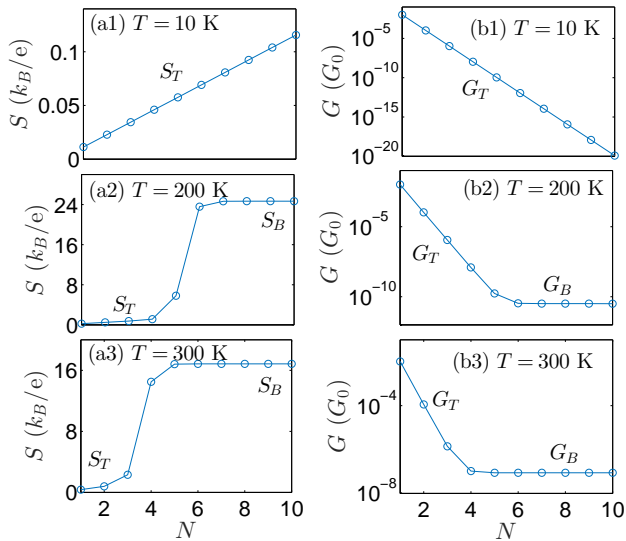}}}
\caption{Coherent transport in the 1D Model.
(a1-a3) Thermopower and (b1-b3) electrical conductance as a function of molecular length
at temperatures (a1,b1) 10 K, (a2,b2) 200 K and (a3,b3) 300 K.
We use $\epsilon_B=0.5$, $v_a=0.05$, $\gamma_{L,R}=0.05$, all in eV.
The different limits, deep tunneling and ballistic motion, are labeled.
}
\label{Fig1}
\end{figure}

\begin{figure*}[htbp]
\vspace{0mm} \hspace{3mm}
{\hbox{\epsfxsize=185mm \epsffile{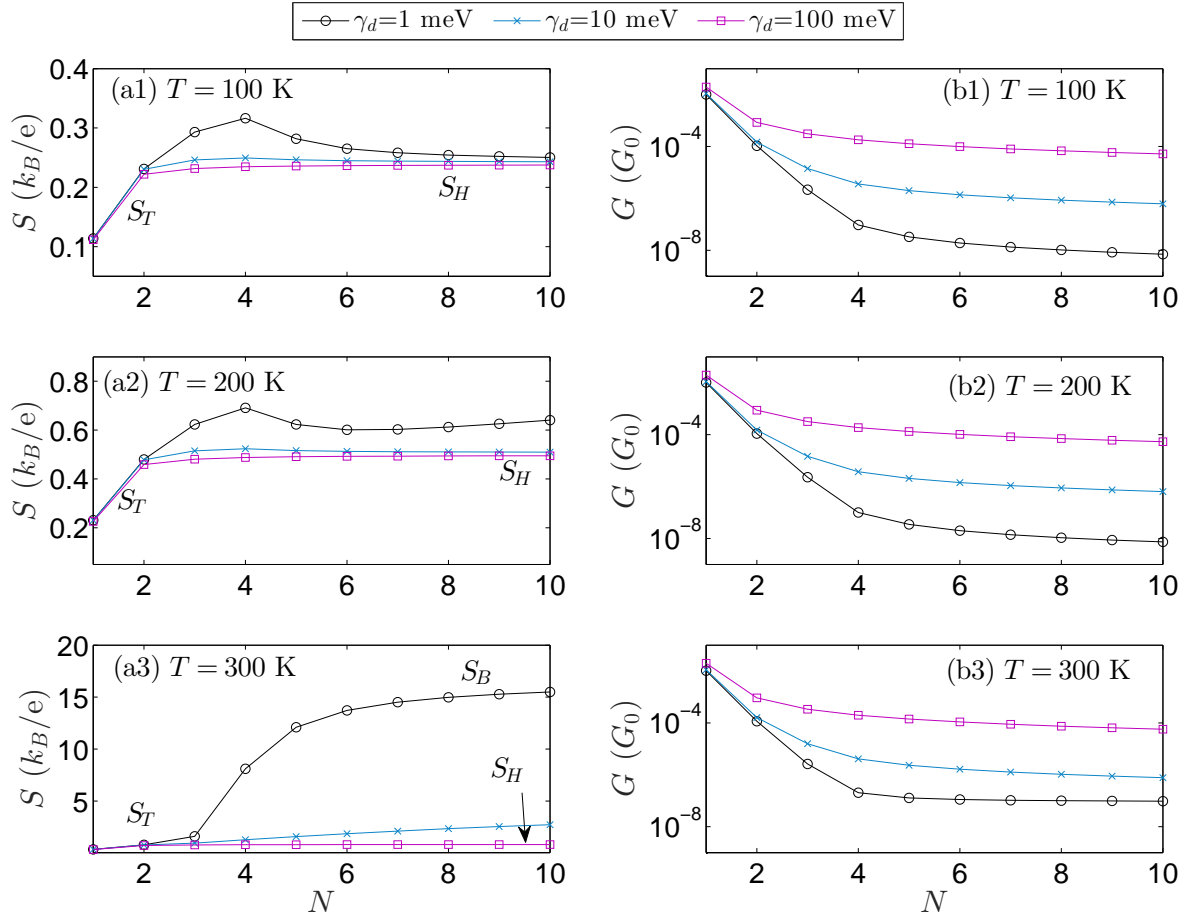}}}
\caption{1D Model with environmental effects.
(a1-a3) Thermopower and (b1-b3) conductance as a function of molecular length 
for the temperatures 100 K, 200 K  and  300 K,
with $\gamma_{d}$=1, 10, 100 meV,
as indicated in the panels. Other parameters are $\epsilon_B=0.5$, $v_a=0.05$, $\gamma_{L,R}=0.05$, in eV.
The different limits for $S$, deep tunneling,  ballistic motion, and hopping  conduction, are marked.
}
\label{Fig2}
\end{figure*}

\begin{figure*}[ht]
\vspace{0mm} \hspace{3mm}
{\hbox{\epsfxsize=185mm \epsffile{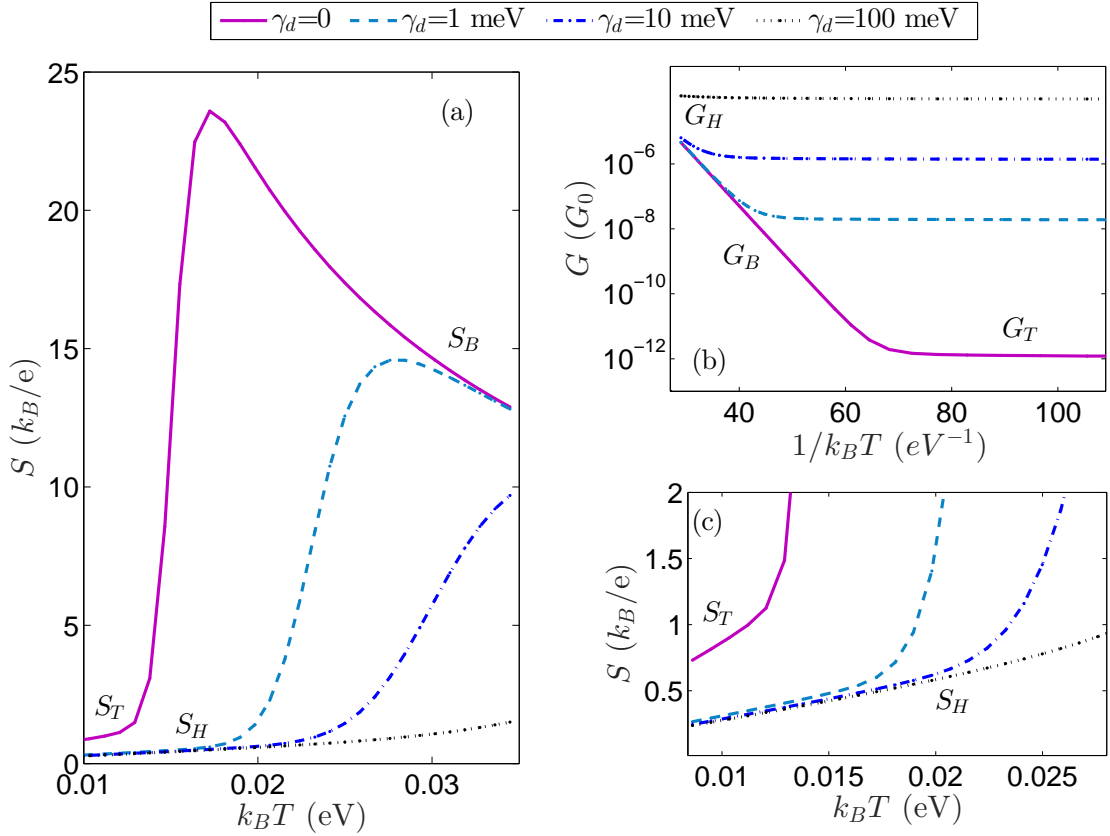}}}
\caption{
(a) Thermopower and (b) electrical conductance as a function of temperature in the 1D model with $N=6$ sites,
$\gamma_d=0,1,10,100$ meV. 
Panel (c) zooms over $S$ in the range 100 - 330 K.
Other parameters are $\epsilon_B=0.5$, $v_a=0.05$, $\gamma_{L,R}=0.05$ all in eV.
We identify by $G_{T,B,H}$ and $S_{T,B,H}$ the limiting  behavior of
 off-resonance  tunneling, on-resonance ballistic, and hopping conduction.
}
\label{Fig3}
\end{figure*}

\begin{figure}[htbp]
\vspace{0mm} \hspace{-4mm}
{\hbox{\epsfxsize=90mm \epsffile{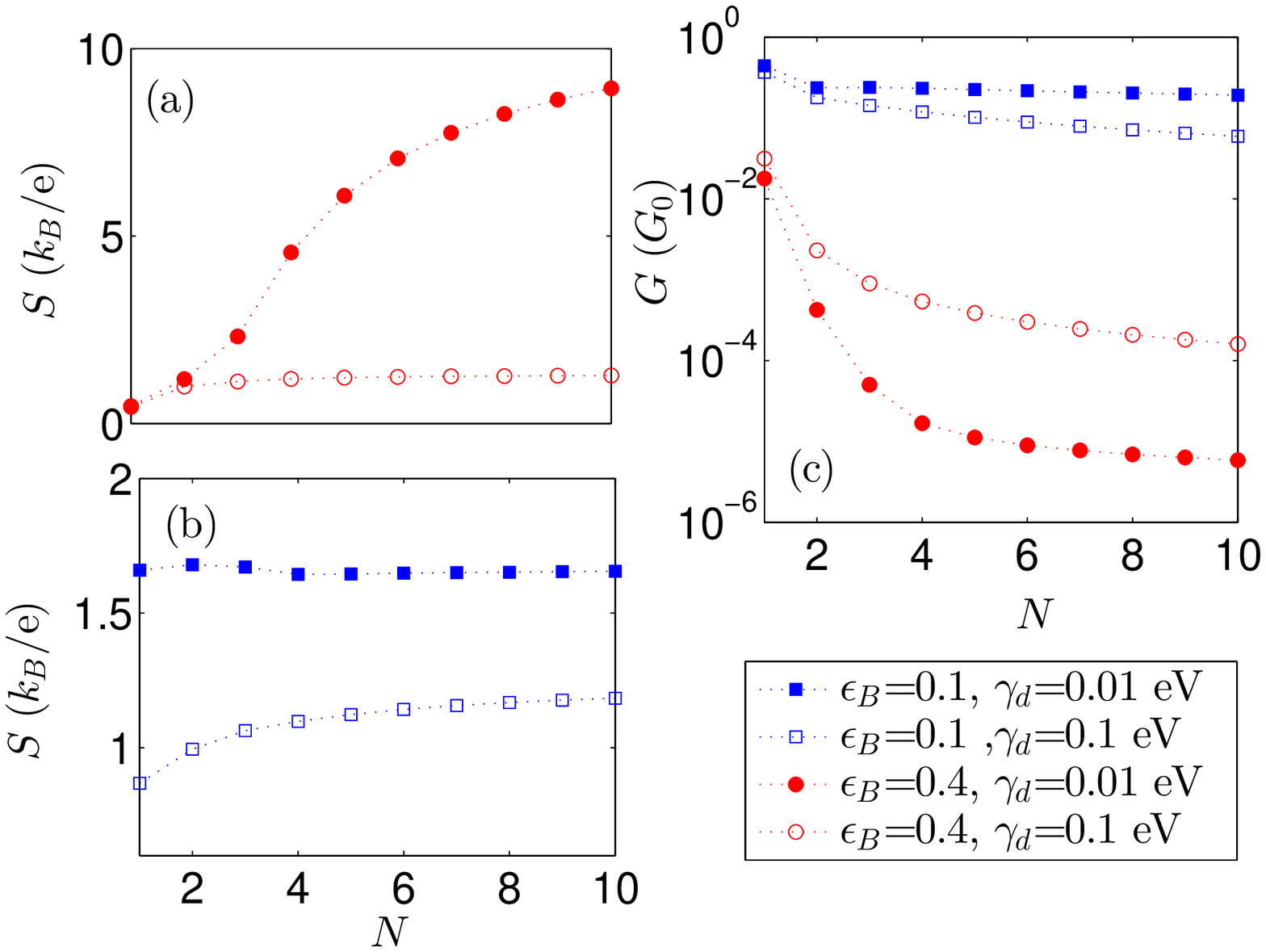}}}
\caption{1D model in the limits $\epsilon_B\gg v_a$ ($\circ$) and $\epsilon_B\sim v_a$ ($\square$).
(a)-(b) Thermopower and (c) electrical conductance as a function of number of sites
at room temperature, $T=300$ K.
Other parameters are $v_a=0.05$, $\gamma_{L,R}=0.05$ all in eV.
}
\label{Fig4}
\end{figure}

\subsection{Simulations}

We study the coherent limit ($\gamma_d=0$) in Fig. \ref{Fig1}. To identify
the tunneling-to-ballistic transition, we follow 
both the electrical conductance and the thermopower.
In the off-resonance tunneling regime 
(low temperature and short junctions), the conductance decays exponentially with
length. At the same time, 
the thermopower grows linearly with $N$, with a slope in precise agreement with Eq. (\ref{eq:STT}).
At high temperatures and for long enough chains, the ballistic regime dominates and 
both the conductance and the thermopower become independent of length, with Eq. (\ref{eq:SB}) 
very accurately predicting the asymptotic value for $S$. 
Specifically, we immediately confirm that in the ballistic regime $S$ scales with $T^{-1}$, 
in a sharp contrast to the off-resonant tunnelling behavior, supporting $S\propto T$.

Figure \ref{Fig2} illustrates the result of incoherent effects on $S$, using $\gamma_d$=1, 10, 100  meV.
A tunneling-to-hopping crossover is observed in both conductance and thermopower if the system is 
maintained at relatively low temperatures and $\gamma_d$ 
is made large enough to support hopping transport. 
%
Specifically, in panel (a1), we use $T=100$ K.
$S$ then grows linearly for short $N$ and saturates to the hopping value 
$S\sim 0.25$ $k_B/e$, independent of $N$ and $\gamma_d$. This number
excellently agrees with the prediction of Eq. (\ref{eq:SH}).
At high temperatures (room temperature in our case), the tail of the Fermi function 
enhances the ballistic contribution---even beyond hopping conduction \cite{Kilgour1}.
Indeed, in panel (a2)  with $T=$200 K we observe that the ballistic component 
is enhanced over the hopping contribution at small $\gamma_d$. 
Finally, in panel (a3) we clearly see that at room temperature, 
the thermopower at $\gamma_d=1$  meV 
follows a ballistic  trend, while at $\gamma_d=100$ meV it displays hopping characteristics.
Simultaneously, we confirm that the electrical conductance displays corresponding transitions between 
different mechanisms: exponential decay with length for short chains, 
length independence when $\gamma_d$ is small and the chain is long (ballistic conduction),
and an Ohmic behavior, $G_H\propto 1/N$, for large $\gamma_d$ in long enough chains.

Activated conduction takes place in both ballistic and hopping regimes.
Therefore, identifying an Arrhenius factor in the conductance  does not conclusively identify
thermally activated incoherent hopping processes \cite{Zant,Nichols}.
This aspect is demonstrated in Fig. \ref{Fig3}(a), with the conductance showing
a typical crossover from an activation-less to an activated form as we increase the temperature, 
even when environmental effects are turned off completely with $\gamma_d=0$, see also Ref. \cite{Kilgour1}. 
Temperature-dependent conductance measurements can  thus be quite confusing to interpret.
In contrast, the temperature dependence of the thermopower clearly separates 
ballistic and hopping contributions:
$S\propto T$ in both the deep tunneling and the hopping regimes, but
$S\propto 1/T$, when electrons ballistically cross the junction, see
Fig. \ref{Fig3}. 
We particularly emphasize (panel c) that around room temperature hopping is the dominant transport mechanism,
when $\gamma_d\sim 100$ meV. 
We also note that at low temperatures $S$ is always linear; the slope corresponds to the molecular length
in the deep tunneling regime, when Eq. (\ref{eq:STT}) is satisfied.

In Fig. \ref{Fig4} we turn our attention to junctions with a low energy barrier, $\epsilon_B\sim v_a$, $k_BT$.
In this limit, the conductance {\it decreases} with increasing $\gamma_d$ (squares),
as opposed to the case with $|\epsilon_B /v_a|\gg 1$ (circles).
We observe that the thermopower weakly depends
on distance when $|\epsilon_B/ v_a|\sim 1$, and that its magnitude is reduced with $\gamma_d$. 
These effects both come about because as we increase $\gamma_d$ we broaden molecular resonances, 
resulting in dampening of on-resonance conduction, and
flattening the transmission function, the derivative of which roughly dictates $S$.

We now comment on the sign of the thermopower in our simulations.
Calculations in this Section assumed electron transport.
The factor $k_B/e$ should be evaluated with the negative sign for the unit of charge, 
to receive negative values for $S$ in  Figs. \ref{Fig1}-\ref{Fig4}.
Since our 1D model concerns LUMO transport, 
negative value for $S$ conforms with the literature \cite{AgraitRev}.

\section{Tunneling to hopping transition in DNA}
\label{DNA}

In the previous section we demonstrated that 
the thermopower exhibits signatures of underlying transport mechanisms. 
Particularly, panel (a1) of Fig. \ref{Fig2} manifests
the characteristic tunneling-to-hopping crossover. 
This result qualitatively captures recent
experimental reports over a related behavior in DNA molecules \cite{Tao16}.  
While in Sec. \ref{1D} we employed a simple 1D tight-binding model to describe generic molecular chains,
our objective in this section is to use a more detailed model Hamiltonian for simulating charge transport in double-stranded 
(ds) DNA structures. Particularly, we consider here 
G:C rich sequences with an inserted A:T block, as well as alternating GC sequences.
See Table I for the list of sequences examined.
We compute simultaneously the conductance and thermopower in these molecules,
with an effort to reproduce and explain recent measurements \cite{Tao16}. 
Our analysis further furnishes us with a better understanding of the limitations of the LBP method in simulating
biomolecules.

Simulating charge transport in DNA molecules is an exceptionally challenging task
given their complex electronic structure and with rich 
structural, solvent and counterion dynamics.
Among the approaches devised to follow charge transport in these systems we recount here
kinetic rate equations \cite{Bixon,Burin,danny}, which capture hopping transport, and
the Landauer approach, which describes only the coherent limit  \cite{cuniberti10}, see a recent application
in Ref. \cite{dubi}.
Simulating intermediate coherent dynamics in DNA is an intricate task:
This regime can be followed with phenomenological tools \cite{Wolf,probeQi,spin,William}, or by 
combining classical molecular dynamics simulations (for describing the effect of backbone, solvent, counterions,
and he DNA internal structural fluctuations)
with quantum mechanics/molecular mechanics methodologies \cite{beratanRev, cunibertiJCP09,cunibertiLee,Gutierrrez1,cunibertiNJP10}.
Alternatively, in Ref. \cite{Beratan16}
charge transport in DNA was simulated using a stochastic Schr\"odinger equation while taking into account
temporal and spatial fluctuations in  electronic parameters.
In other methods, one explicitly includes the interaction of transport charges with
selected DNA vibrational modes using e.g. Green's function approaches \cite{cunibertiphonon},
quantum rate equations \cite{PeskinPCCP}, or semiclassical approximations \cite{BerlinSC}.

Our simulations here are based on the LBP method as explained in Sec. \ref{Method}.
In a recent application of this technique to DNA \cite{William},
we demonstrated that the LBP method can yield results in qualitative (and sometimes quantitative)
agreement with experiments, and uncovered an intermediate coherent-incoherent transport regime \cite{Ratner15}.
An important caveat of our simulations is the fact that 
ballistic transport is promoted beyond what we expect in reality,
since the underlying electronic structure used is static. 
To overcome this deficiency, and limit the ballistic conduction, in our simulations below we
have reduced the temperature while keeping environmental effects active with a finite $\gamma_d$ of 5-50 meV.
This modification allows us to nicely recover experimental trends
for both the electrical conductance and the thermopower, see Figs. \ref{DNA1}-\ref{DNA2}.
Note that in our simulations the parameter $\gamma_d$, which controls the rate of electron
 scattering with environmental degrees of freedom, is independent from the temperature.
The single role of the temperature is therefore to broaden the Fermi functions, 
thus directly controlling the injection of ballistic charges at the tail of the Fermi function.

\begin{figure}[htbp]
\vspace{0mm}
{\hbox{\epsfxsize=85mm \epsffile{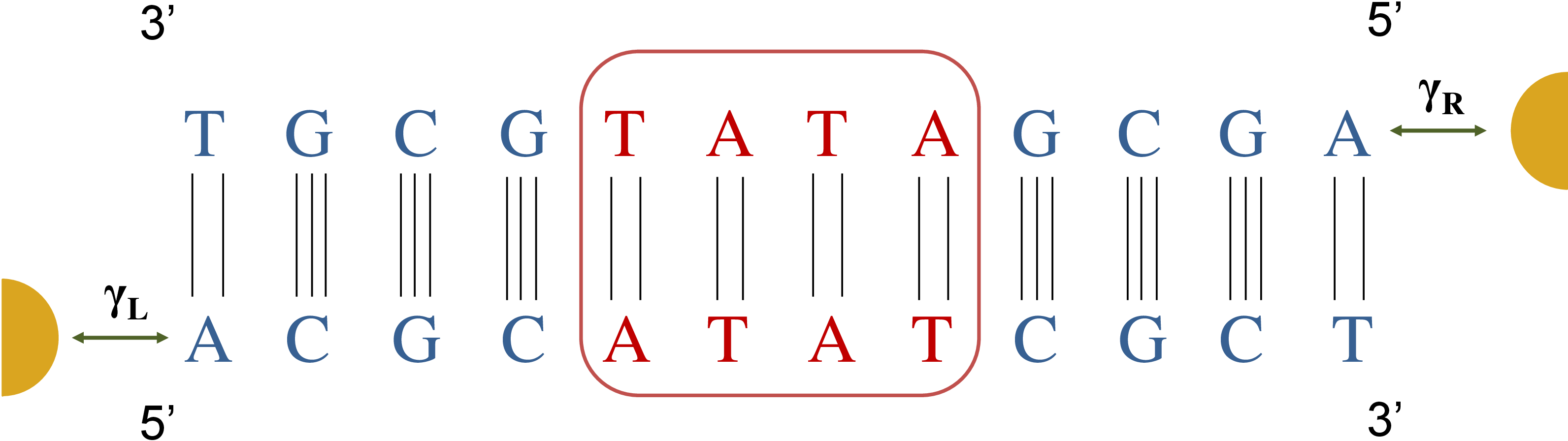}}}
\caption{Example for a 12-base pair DNA duplex simulated in this work, ACGC(AT)$_2$GCGT.
The AT barrier is highlighted by the box.
}
\label{Figs2}
\end{figure}

\vspace{5mm}
\begin{table*}[htbp]
\hspace{10.4mm } Table I:   {\bf  Sequences studied in this work, with measurements reported in Ref. \cite{Tao16}. }\\

\begin{tabular}{c c c }
\hline
\hline
 number of base pairs   \vline &  5'-A(CG)$_m$T-3'  \hspace{8mm }    & 5'-ACGC(AT)$_m$GCGT-3' and \\ [0.5ex]
\hspace{2.4mm }  (molecular length) \vline &   $m=3-8$  \hspace{1mm}  &  5'-ACGC(AT)$_{m-1}$AGCGT-3', $m=1-4$\\ [0.5ex]
\hline
\hspace{12mm }8 \hspace{15mm} \vline &  A(CG)$_3$T &    \\
\hline
 \hspace{11mm } 9 \hspace{14.9mm} \vline &  &  ACGCAGCGT \\
\hline
\hspace{10mm } 10 \hspace{14.3mm} \vline & A(CG)$_4$T  &  ACGCATGCGT\\
\hline
\hspace{10mm } 11 \hspace{14.3mm} \vline &  &    ACGCATAGCGT \\
\hline
\hspace{10mm } 12 \hspace{14.3mm} \vline &  A(CG)$_5$T   &    ACGC(AT)$_2$GCGT \\
\hline
\hspace{10.7mm }13 \hspace{14.7mm} \vline &  &    ACGC(AT)$_2$AGCGT \\
\hline
\hspace{10.7mm }14 \hspace{14.7mm} \vline &   A(CG)$_6$T  &    ACGC(AT)$_3$GCGT \\
\hline
\hspace{10.5mm }15 \hspace{14.7mm} \vline &  &    ACGC(AT)$_3$AGCGT \\
\hline
\hspace{10.5mm }16 \hspace{14.7mm} \vline &   A(CG)$_7$T  &   ACGC(AT)$_4$GCGT \\
\hline
\hspace{10.5mm }18 \hspace{14.7mm} \vline &    A(CG)$_8$T &     \\
\hline
\hline
\end{tabular}
\label{table:seq}
\end{table*}



\subsection{Model}

We model charge transport in DNA 
using a tight-binding ladder model Hamiltonian, see e.g. Refs. \cite{cunibertiphonon, ladder1,ladder2,ladder3,Wolf}.
This Hamiltonian describes the topology of a ds-DNA molecule which is $n$ base-pairs long,
with each site representing a particular base; $N=2n$ is the total number of bases.
We assume that charge transport takes place along the base-pair stacking and we ignore the backbone,
\bea
\hat H_M&=&\sum_{j=1}^n \Bigg[
\sum_{s=1,2} \epsilon_{j,s}\hat c_{j,s}^{\dagger}\hat c_{j,s} 
+ \sum_{s\neq s'=1,2} t_{j,ss'}\hat c_{j,s}^{\dagger}\hat c_{j,s'}
\nonumber\\
&+&
 \sum_{s,s'=1,2} t_{j,j+1,ss'}(\hat c_{j,s}^{\dagger}\hat c_{j+1,s'} + h.c.) \Bigg].
\eea
%
The index $s=1,2$ identifies the strand. $\hat c_{j,s}^{\dagger}$ creates a hole on strand $s$ 
at the $j$th site with an on-site energy $\epsilon_{j,s}$. $t_{j,ss'}$ and $t_{j,j+1,ss'}$
are the electronic matrix elements between nearest neighboring bases.
This model mimics the topology of the ds-DNA molecule; helical effects are 
taken into account within renormalized electronic parameters.

We use the parametrization of Ref. \cite{BerlinJacs}, 
distinguishing between backbone orientations (5' and 3').
In this parametrization, on-site energies depend on the identity of neighboring bases.
Here, following Ref. \cite{Wolf}, we simplify this description and assign a single value (averaged)
for on-site energies for each base, see Table II. 
Electronic matrix elements were taken directly from Ref. \cite{BerlinJacs}.


We connect the DNA molecule to metal leads as sketched in Fig. \ref{Figs2}, 
oriented so as to correspond to experiment \cite{Tao16}.
Environmental effects (structural motion, solvent, counterions) are captured by the probes' condition
implementing phase loss and inelastic effects.

Beyond the molecular electronic structure of the ds-DNA,
three additional parameters should be provided as input to LBP equations:
The position of the Fermi energy $\epsilon_F$ relative to the molecular states,
the metal-molecule hybridization energy $\gamma_{L,R}$, and the electron-environment interaction energy,
encapsulated within the parameter $\gamma_d$.
In principle, we could use a range of values for $\gamma_d$
to capture the susceptibility of different bases
and sites along the DNA to environmental interactions.
Here, for simplicity, we use a single value
 $\gamma_d$, identical for all bases and sites.
Furthermore, we note that while one could carefully optimize these three parameters, 
$\epsilon_F$, $\gamma_{L,R}$ and $\gamma_d$, to optimally reproduce
experimental results, 
our goal here has been to reach a general understanding 
over the behavior of the thermopower under different mechanisms.
Therefore, we present our results with representative  parameters, reasonable for a junction geometry.

We now point out that by convention, our transport expressions are all written for electrons:
We use the Fermi function $f(\epsilon)$ in the Landauer formula, 
standing for the occupation factor of electrons. However, the
DNA parametrization which we use concerns holes \cite{BerlinJacs}. 
While the conductance is unaffected by the identify of the charge carriers 
(it scales with $e^2$), 
this fact was taken into account in the evaluation of the thermopower, in physical units, as presented in Figs. \ref{DNA1}-\ref{DNA3}.
Note that in simulations presented below, the conductance was calculated including
 both spin species, using $G_0=2e^2/h$.


\vspace{4mm}
\label{tab:title}
\begin{center}
Table II: Selected on-site energies and inter-strand electronic coupling (eV).
(See \cite{BerlinJacs} for full parameter set.)\\ 
\begin{tabularx}{.45\textwidth} { c c  c  c c c  }
\hline
\hline
$\epsilon_G$ \hspace{3mm} & $\epsilon_A$ \hspace{3mm} & $\epsilon_C$ \hspace{3mm} & $\epsilon_T$ \hspace{3mm} & $t_{\rm{G||C}}$ \hspace{3mm} &  $t_{\rm{A||T}}$\\
\hline
8.178 \hspace{3mm} & 8.631 \hspace{3mm} & 9.722 \hspace{3mm}& 9.464 \hspace{3mm} & -0.055 \hspace{3mm}& -0.047 \\
\hline
\hline
\end{tabularx}\par
\end{center}
\vspace{4mm}


\begin{figure*}[ht]
\vspace{0mm} \hspace{-4mm}
{\hbox{\epsfxsize=180mm \epsffile{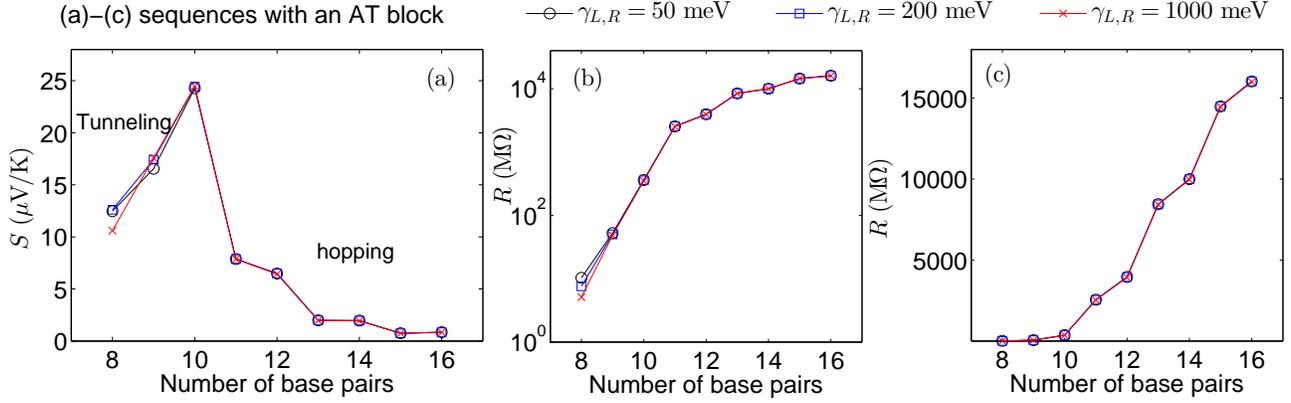}}}
\caption{
(a) Thermopower and (b)-(c) electrical resistance in logarithmic and linear scale
of DNA sequences with an AT block, ACGC(AT)$_m$GCGT and  ACGC(AT)$_{m-1}$AGCG ($m=1-4$), see Table I.
Simulations were performed at $T$=5 K to attenuate the contribution of ballistic electrons.
Other parameters are  $\gamma_{d}=10$ meV, $\gamma_{L,R}$=50, 200, and 1000 meV. The
Fermi energy is placed at the energy of the G base, $\epsilon_F=\epsilon_G$.
}
\label{DNA1}
\end{figure*}

\begin{figure*}[ht]
\vspace{0mm} \hspace{-4mm}
{\hbox{\epsfxsize=180mm \epsffile{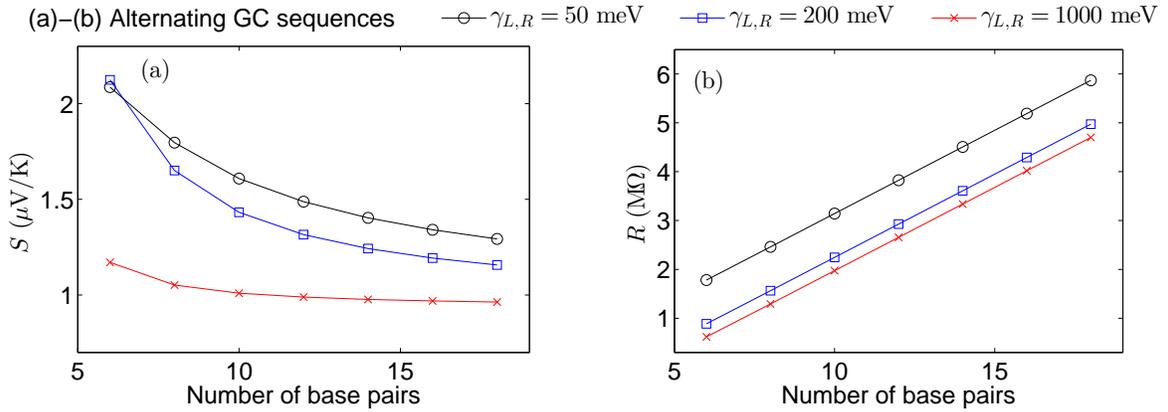}}}
\caption{ 
(a) Thermopower and (b) electrical resistance of A(CG)$_m$T sequences, see Table I. 
Simulations were performed at $T$=5 K,  $\gamma_{d}=50$ meV,  $\epsilon_F=\epsilon_G$.
}
\label{DNA2}
\end{figure*}


\subsection{Results}

The Seebeck coefficient and the resistance of DNA sequences with an AT block are included in Fig. \ref{DNA1}, 
manifesting a clear  tunneling-to-hopping transition around 11 base pairs. 
The tunneling behavior is characterized by an exponential enhancement of resistance and by 
a linear increase of the thermopower. 
The hopping regime shows an Ohmic resistance, and a marginal (yet nonzero) thermopower, below the coherent value.
These qualitative trends are in an excellent agreement with the experiment \cite{Tao16}.
Our simulations were performed at low temperatures to cut down on ballistic electrons which contribute 
beyond what is expected in reality, where static and dynamic disorder naturally reduce ballistic conduction.
We further confirmed that our results only weakly depend on the
metal-molecule hybridization energy.
Note that we were not able to reproduce measurements in a quantitative manner, 
neither for the resistance, nor for the thermopower.

We recall that our expressions for linear chains predict that $S_T/(2n)=S_H/4$, see Eqs. (\ref{eq:STT}) and (\ref{eq:SH}); 
in the ds-DNA the number of base pairs $n$ corresponds to the length.
We examine the reported results of Ref. \cite{Tao16}: The tunneling value is   $S_T(n=10)=6$ $\mu$V/K, 
while $S_H(n=16)=2$ $\mu$V/K. The ratio between these 
values reasonably satisfy theoretical predictions.

The Seebeck coefficient and the resistance of DNA sequences with alternating GC base-pairs (see Table I) are displayed in
Fig. \ref{DNA2}. Here, as confirmed in Ref. \cite{William}, the resistance increases linearly with length throughout,
in accord with a multi-state hopping conduction mechanism.
The thermopower shows monotonic behavior, continuously decreasing with length.
We emphasize that (\ref{eq:SH}) is invalid in the present conditions, as it was derived assuming high energy 
barriers relative to temperature
These simulations were also performed at $T=5$ K, as explained above.
We adopted here the value $\gamma_d=50$ meV, to reproduce experimental values for resistance.
It should be noted however, as we show in Fig. \ref{DNA4}, that our results for this family of sequences are quite robust in the range $\gamma_d=5-100$ meV.

\begin{figure*}[ht]
\vspace{0mm} \hspace{-4mm}
{\hbox{\epsfxsize=170mm \epsffile{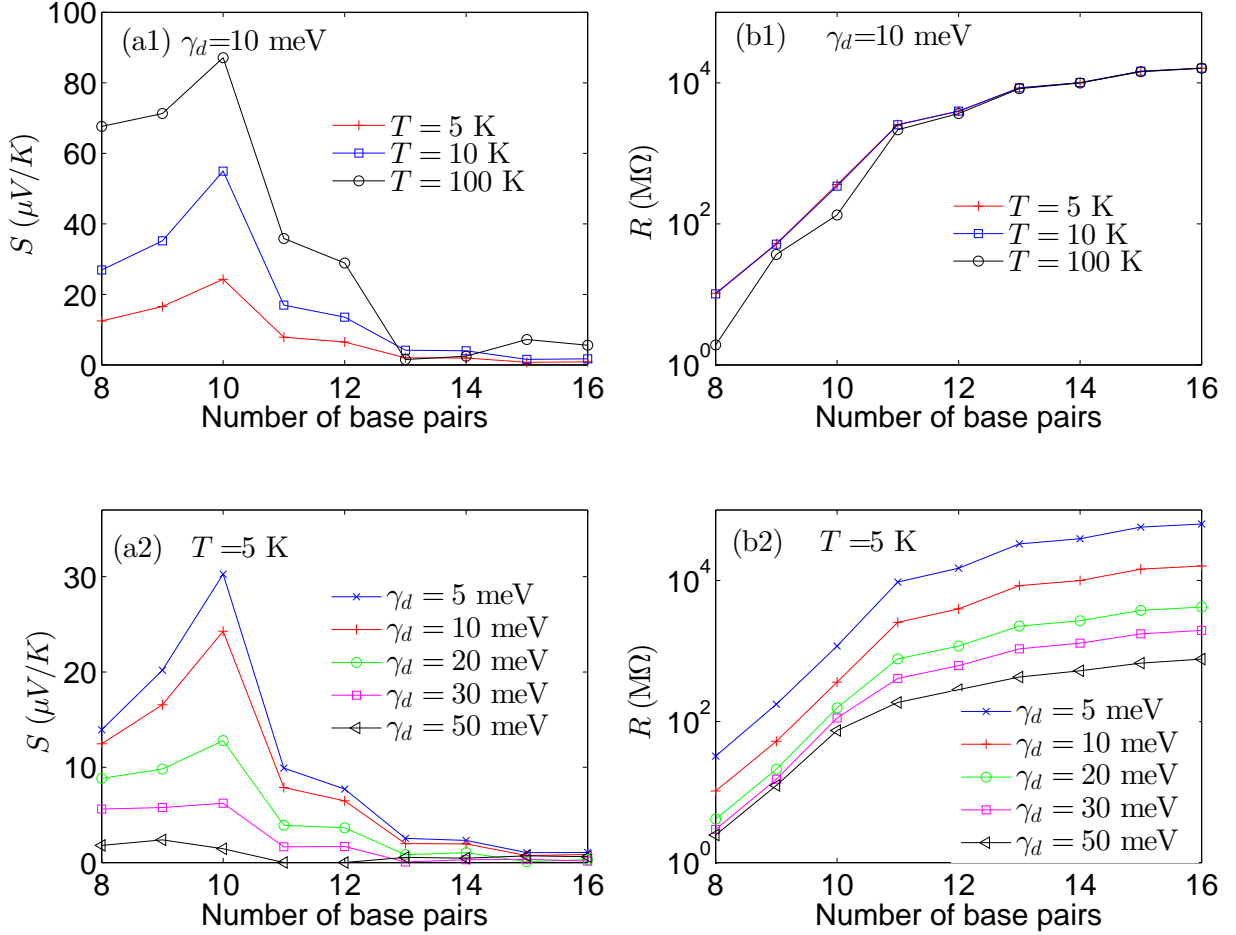}}}
\caption{
(a) Thermopower and (b) electrical resistance 
of DNA sequences with an AT block, ACGC(AT)$_m$GCGT and  ACGC(AT)$_{m-1}$AGCG ($m=1-4$), see Table I,
for a range of temperatures and environmental interactions, $\gamma_d$.
Other parameters are  $\gamma_{L,R}=50$ meV,  $\epsilon_F=\epsilon_G$.
}
\label{DNA3}
\end{figure*}

\begin{figure*}[ht]
\vspace{0mm} \hspace{-4mm}
{\hbox{\epsfxsize=170mm \epsffile{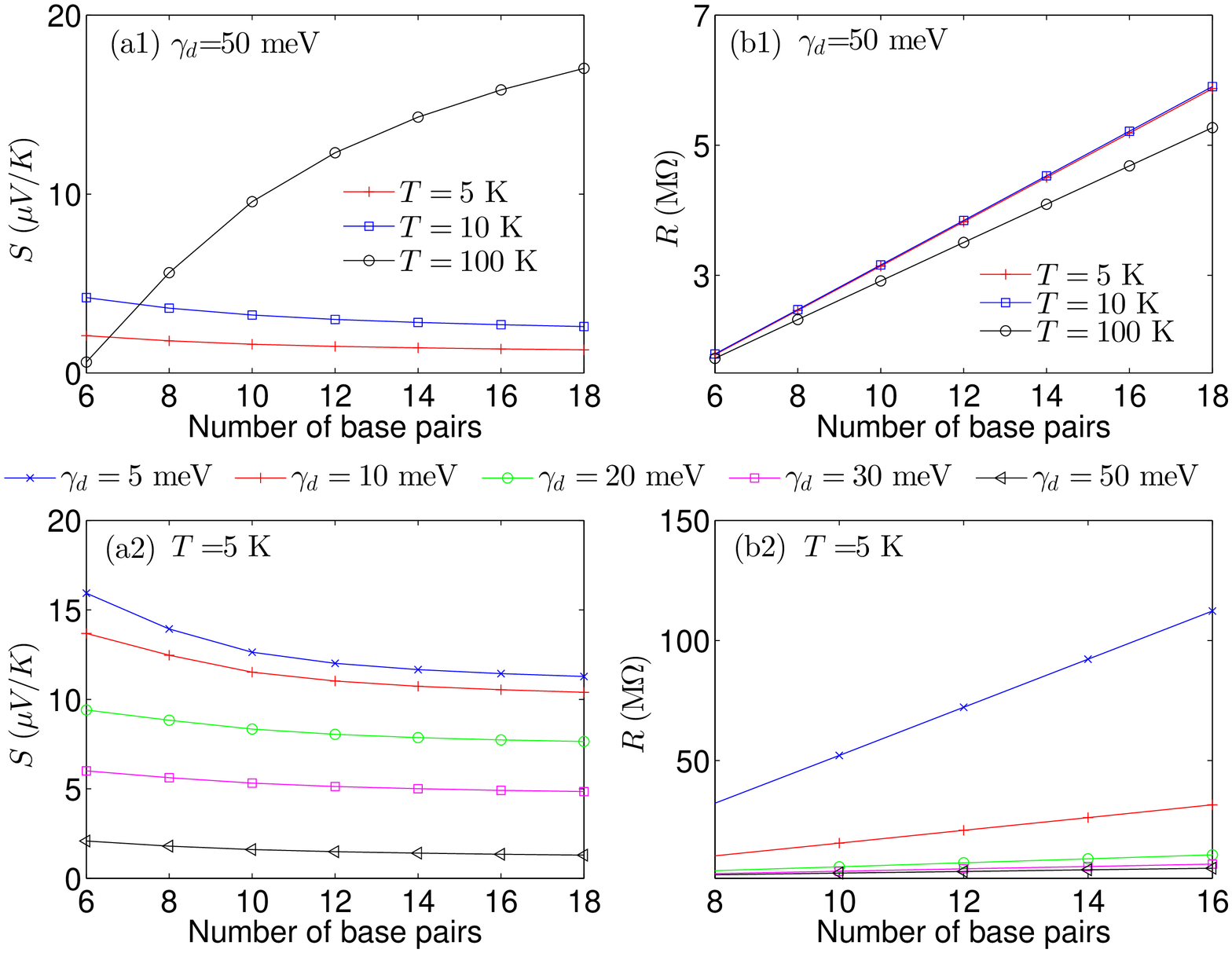}}}
\caption{
(a) Thermopower and (b) electrical resistance
of (CG)$_m$ sequences, see Table I,
for a range of temperatures and environmental interactions $\gamma_d$.
Other parameters are  $\gamma_{L,R}=50$ meV, $\epsilon_F=\epsilon_G$.
}
\label{DNA4}
\end{figure*}

To establish our observations,
in Figs. \ref{DNA3}-\ref{DNA4} we comprehensively test the behavior
of the DNA junctions at different temperatures and values of $\gamma_d$. 
We do not display results with different values of the Fermi energy, as we observed that
varying it by $\pm 0.1$ eV lead to a significant increase of the resistance, 
further away from  experimental values. 
Fig. \ref{DNA3} demonstrates that in sequences with an A:T barrier,
the tunneling and hopping regimes are established when the temperature is rather low $T=5-100$ K 
and $\gamma_d=5-20$ meV.  Recall that we perform  here low-temperature simulations so as to limit the injection of 
ballistic electrons, which are suppressed in real systems by dynamics and static fluctuations.
Fig. \ref{DNA4}(a1) specifically shows that at 100 K, $S$ grows  due to ballistic conduction. 
It is important to note that at the same time, the resistance does not well evince on the ballistic contribution 
(see panel b1) as it is over-dominated by the Ohmic behavior.

Our simulations indicate that a simultaneous examination of the electrical conductance 
and the thermopower is highly beneficial
for correctly identifying transport mechanisms. 
In some situations, the resistance shows a linear enhancement, indicating on hopping conduction,
yet at the same time the thermopower pinpoints on a significant contribution 
of on-resonance electrons.
Regarding the experiment reported in Ref. \cite{Tao16}, 
our VTP simulations support the identification of the different transport limits.
We also suggest measuring the temperature dependence of $G$ and $S$, around room temperature, for
this family of DNA sequences (Table I).
Establishing the relationship  $S \propto T$ in both short and long sequences 
would further support the claim of hopping-dominated conduction in long duplexes.

\section{Summary}
\label{Summ}
We developed analytical expressions for the thermopower 
of single molecule junctions, covering three different transport regimes:
(i) off resonant (superexchange) and (ii) on-resonant (ballistic) tunneling, as well 
as (iii) multi-step hopping.
Using the Landauer-B\"uttiker probe technique, 
we simulated the conductance and thermopower of 1D chains
and conclusively identified transport mechanisms with varying temperature and molecular length. 
While the distance dependence of the thermopower can separate tunneling behavior from longer-range 
ballistic and hopping mechanisms,
the temperature dependence is necessary to conclusively distinguish between these two, 
which are both largely distance independent.

We applied the probe approach to simulate the resistance and thermopower in two families of DNA molecules: 
sequences with an AT barrier which display a tunneling-to-hopping crossover with increasing barrier length, 
and alternating GC sequences, supporting hopping transport.
Our LBP simulations qualitatively agree with recent measurements \cite{Tao16}, 
providing a critical theoretical affirmation of these results. 
Specifically, we were able to reproduce the tunneling-to-hopping crossover close to the correct position.

Simulations of DNA structures were performed at rather low temperature
to minimize the contribution of ballistic electrons 
(while keeping environmental effects with a finite value for $\gamma_d$). 
In reality, ballistic charge conduction is inhibited in DNA given its complex electronic structure and rich nuclear motion.
Future work will incorporate energetic disorder with the aim of mimicking these rich dynamics,
and naturally attenuating on-resonance coherent conduction.

This study complements our recent work on the LBP method, demonstrating its adequacy for simulating 
transport phenomena in single molecule junctions \cite{Kilgour1, Kilgour2, Kilgour3, William}.
A concrete comparison between LBP simulations and other techniques, where 
many-body electron-phonon effects are explicitly included, is still missing,
and will be the focus of future work.


\begin{acknowledgments}
The work was supported by the Natural Sciences and Engineering Research Council of Canada and 
the Canada Research Chair Program.
The work of Michael Kilgour was partially funded by an Ontario Graduate Scholarship. 
Roman Korol was partially funded by the University of Toronto Excellence Research Fund.
\end{acknowledgments}

{}

\end{document}